\documentclass[journal]{IEEEtran}
\usepackage{cite}
\usepackage{xcolor}
\usepackage{lineno}
\usepackage{subcaption}
\usepackage[hidelinks]{hyperref}
\usepackage{balance}
\usepackage{booktabs}
\usepackage{multirow}
\usepackage{orcidlink}
\usepackage{tabularx}
\usepackage{array}
\newcolumntype{C}{>{\centering\arraybackslash}X}
\usepackage{graphicx}
\begin{document}
%
\title{RF-Resolved RIN Engineering of High-Power CW Yb Fiber Laser Architectures Using Seed Topology, Seed Power, Pumping Geometry and Cavity FSR}
%
%
%
\author{
Anitta Augustine\orcidlink{0009-0002-3206-9945},
V.~Meher\orcidlink{0000-0006-4217-0608},
Shantanu Dambare\orcidlink{0009-0006-7228-8922},
and V.~Balaswamy\orcidlink{0000-0002-7732-3452}%
\thanks{This work was supported in part by the Anusandhan National Research Foundation (ANRF) under Grant SRG/2023/001399 and Grant EEQ/2023/000570, and in part by the ISRO Space Technology Cell, Indian Institute of Science, Bengaluru, under Grant ISTC/EEC/BV/493.}%
\thanks{The authors are with the Department of Electrical Communication Engineering (ECE), Indian Institute of Science, Bengaluru, Karnataka 560012, India. Corresponding author: V.~Balaswamy (e-mail: velpulab@iisc.ac.in).}
}



%
%

\markboth{Journal of Lightwave Technology}%
{Anitta Augustine \MakeLowercase{\textit{et al.}}: RF-Resolved RIN Engineering of CW Yb Fiber Lasers}
%



\maketitle

\begin{abstract}
We report, to the best of our knowledge, the first controlled, like-for-like experimental mapping of how four practical design parameters--seed-cavity topology, seed power, amplifier pumping geometry, and seed-cavity free spectral range (FSR)--jointly shape the radio-frequency (RF)-resolved relative intensity noise (RIN) of high-power continuous-wave (CW) Yb-doped fiber lasers. We compare a standalone bidirectional fiber Bragg grating (FBG) oscillator, an FBG-seeded master-oscillator power amplifier (MOPA), and unidirectional ring-seeded MOPAs with FSRs of $\sim$11.5 and $\sim$5~MHz, all near 1064~nm at a common 72~W output, with RIN measured from 10~kHz to 1~GHz. Across all settings, the low-frequency RIN is technical-noise limited and nearly architecture-independent, whereas RIN above 160~kHz is governed by the seed-cavity topology and its transfer through the amplifier. Higher seed power suppresses the mid- and high-frequency RIN, but the seed-alone and amplified-output comparison reveals opposite transfer for the two topologies: for FBG seeding the 2~MHz--1~GHz RMS RIN increases through amplification (3.66\% to 4.14\% at 15~W seed), whereas for ring seeding it decreases (1.47\% to 1.35\%). Forward pumping gives the best power-scaling/noise trade-off, while reducing the ring FSR from $\sim$11.5 to $\sim$5~MHz preserves the high-frequency RIN advantage, confirming it is not a mode-spacing effect. At matched conditions, ring seeding lowers the integrated RMS RIN by 2--3$\times$ (160~kHz--2~MHz) and $\sim$3$\times$ (2~MHz--1~GHz, $\sim$9--10~dB to near $-127.4$~dBc/Hz). These results yield band-resolved design rules linking each band to its dominant lever. The unidirectional ring-seeded MOPA thus offers a simple, scalable route to broadband pump-RIN suppression for nonlinear and frequency-conversion photonics.

\end{abstract}

\begin{IEEEkeywords}

Relative intensity noise, RF-resolved RIN, high-power CW fiber laser, Yb-doped fiber MOPA, seed-cavity topology, unidirectional ring-cavity seed laser, fiber Bragg grating oscillator, amplifier pumping geometry, cavity free spectral range.

\end{IEEEkeywords}

%
\IEEEpeerreviewmaketitle

\section{Introduction}

High-power continuous-wave (CW) ytterbium (Yb)-doped fiber lasers have become enabling sources for modern photonics because of their high optical efficiency, excellent power scalability, near-diffraction-limited beam quality, moderate optical bandwidth, and robust all-fiber implementation \cite{richardson2010high,dong2016fiber}. Beyond their direct use in material processing, welding, and precision machining \cite{dong2025past}, they are widely used as pump sources for nonlinear and frequency-conversion systems, including cascaded random-Raman fiber lasers \cite{supradeepa2017raman}, solid-state Raman lasers \cite{liu2023high}, supercontinuum (SC) generation \cite{travers2010continuous}, harmonic generation \cite{politko2013second,han2025}, and optical parametric devices \cite{das2009broadband}. In these systems, CW Yb-doped pump fiber lasers are not merely optical-power sources, as the nonlinear conversion responds not only to average power but also to rapid pump-power fluctuations. These fluctuations are transferred to the nonlinear frequencies generated, and therefore the relative intensity noise (RIN)--and specifically its distribution across radio frequencies (RF)--directly governs the conversion efficiency, spectral purity, temporal stability, and coherence of the output.

The required noise behavior, however, is application-dependent. In CW-pumped supercontinuum generation and cascaded random-Raman fiber lasers, low-RIN pumping is beneficial because pump fluctuations can be transferred to newly generated wavelengths and cascaded Stokes orders, thereby degrading continuum stability and Raman spectral purity\cite{dematos2004temporal,babin2005relative}. By using phase-modulated single-frequency amplifiers and filtered amplified-spontaneous-emission (ASE) sources, which mitigate the relaxation oscillations and longitudinal-mode beating commonly associated with conventional multi-longitudinal-mode (MLM) lasers, the spectrally averaged RIN of supercontinua \cite{dematos2004temporal,cheng2022supercontinuum,hudda2024low} and the spectral purity of cascaded Raman lasers \cite{dong2018high,balaswamy2019high,ye2019pump,deheri2022cascaded} have been substantially improved. However, these approaches achieve noise reduction mainly by changing the pump-source class rather than by engineering a given architecture. Further, they typically require additional phase modulation, spectral filtering, isolation, and cascaded amplification stages, increasing system complexity and cost. In second-harmonic generation, by contrast, the random-phase intensity fluctuations of an MLM source have been shown to increase the conversion efficiency relative to a single-frequency pump of equal average power \cite{politko2013second}. Taken together, these results show that the pump architecture strongly affects the performance of the nonlinear frequency conversion systems and that the design objective is not a universally low RIN but a noise spectrum matched, across the relevant RF bands, to the target process. This raises a concrete design question: \emph{given a required high-power CW Yb pump source, how should one choose and operate the laser architecture to obtain a desired RF-resolved RIN spectrum?}

Despite its practical importance, this question remains largely unanswered, and existing RIN studies have approached the problem from related but distinct directions. Studies of Yb-doped single-frequency and ultrafast amplifier chains have measured RIN transfer functions and established that seed-noise and pump-noise pathways exhibit different frequency responses, with amplifier saturation, gain dynamics, and final-stage pump noise playing important roles \cite{guiraud2016high,jauregui2017optimizing,gierschke2019relative,liu2023intensity}. The fundamental intensity-noise floor of such amplifiers has been benchmarked against the quantum limit \cite{popp2020quantum}. In addition, active feedback control has been used to suppress intensity noise \cite{zhao2017gain}, while pump geometry and gain saturation have been exploited to reduce low-frequency RIN while increasing output power in single-frequency amplifiers \cite{zeng2023simultaneous}. Other studies have compared pump-source classes, such as fiber oscillators, MOPAs, and ASE sources, showing that ASE sources can provide lower RIN because they avoid relaxation oscillations and longitudinal-mode beating \cite{cheng2020relative}. These studies, however, either focus on single-frequency or ultrafast systems, are limited mainly to the Hz-to-MHz range, examine transfer functions rather than complete source architectures, or compare fixed source classes without isolating the role of practical design variables. More fundamentally, across all these systems, the RIN of a pump architecture is treated as a property to be characterized rather than a parameter to be designed. This is especially true for high-power MLM CW Yb fiber lasers, for which a controlled, like-for-like experimental study mapping how the practical design variables — seed-cavity topology, seed power, amplifier pumping geometry, and cavity free spectral range (FSR) — jointly shape the RF-resolved RIN from kHz to GHz is still lacking.

This gap matters because different RF bands originate from different physical and technical contributions and therefore respond to different design choices. Low-frequency RIN is primarily associated with pump-current fluctuations, thermal effects, environmental perturbations, and slow gain variations; intermediate-frequency RIN is typically shaped by relaxation oscillations, gain-mediated pump-to-signal transfer, and amplifier saturation; and high-frequency RIN is generally influenced by the seed longitudinal-mode structure, mode beating, cavity FSR, feedback, and amplifier-added noise. A single integrated RIN value therefore cannot guide the selection of a pump laser for a Raman cascade, a supercontinuum source, or a frequency-conversion stage. What is needed is an RF-resolved design framework that maps measurable noise bands onto controllable laser-design parameters.

We have recently taken a first step toward such a framework. Using conventional, low-cost MLM architectures, we showed that replacing an FBG-cavity pump with a unidirectional ring-cavity pump lowers the high-RF-band RIN, improving supercontinuum RIN \cite{meher2026wavelength} and cascaded random-Raman spectral purity \cite{dambare2026spectral}; in the latter, the pump RIN was resolved into three RF windows and the ring-cavity advantage was found to persist even when the ring FSR was matched to that of the FBG cavity. These studies established the seed-cavity topology and the cavity FSR as RIN design levers, but only at a single operating point — a fixed seed power and forward pumping — leaving open how the RF-resolved RIN depends on the operating parameters a designer actually sets.

In this work, we provide the complete framework. We experimentally compare a standalone bidirectional FBG oscillator, an FBG-seeded MOPA, and unidirectional ring-cavity-seeded MOPAs with cavity FSRs of approximately 11.5 and 5 MHz, all operating near 1064 nm with $\sim$~2 nm optical bandwidth and delivering up to $\sim$87~W under matched gain-fiber, pump-wavelength, and output-power conditions. The RIN is measured from 10 kHz to 1 GHz and analyzed in three RF windows (10 kHz--159 kHz, 160 kHz--2 MHz, and 2 MHz--1 GHz). Going beyond architecture alone — and beyond a conference precursor that compared the three architectures at a single seed power under forward pumping \cite{augustine2026radio} — we systematically vary the seed power from 5 to 15 W and compare forward, backward, and bidirectional pumping for each architecture, so that every controllable parameter is tied to a specific RF-band response. We find that the low-frequency RIN is weakly architecture-dependent and primarily technical-noise-limited, whereas the RIN above 160 kHz is governed by the seed-cavity topology and its transfer through the amplifier: the FBG-seeded amplifier retains and partly enhances the seed's high-frequency mode-beating noise, whereas the ring-seeded amplifier damps it. Within each architecture, higher seed power and forward pumping consistently yield the lowest mid- and high-frequency RIN, though their influence is secondary to that of the topology; the ring-cavity advantage is robust across all of these conditions and persists when the ring FSR is reduced toward that of the FBG cavity, indicating that it is not simply a longitudinal-mode-spacing effect, although the lengthened cavity also modifies the photon lifetime and gain dynamics. The result is a dependable design recipe — a unidirectional ring-cavity seed, forward pumping, and adequate seed power — for low-RIN operation without active stabilization or added system complexity. To the best of our knowledge, this is the first controlled, like-for-like experimental mapping of how these design variables jointly shape the band-resolved RIN of high-power MLM CW Yb fiber lasers from kHz to GHz. The remainder of the paper is organized as follows: Section II describes the experimental configurations and the RF-resolved RIN measurement procedure; Section III presents the dependence of the band-resolved RIN on architecture, seed power, pumping geometry, and FSR; and Section IV summarizes the resulting design rules.

\section{Experimental Configuration and RIN Measurement}
\label{sec:setup}
\begin{figure*}[t]
\centering
\includegraphics[width=\textwidth]{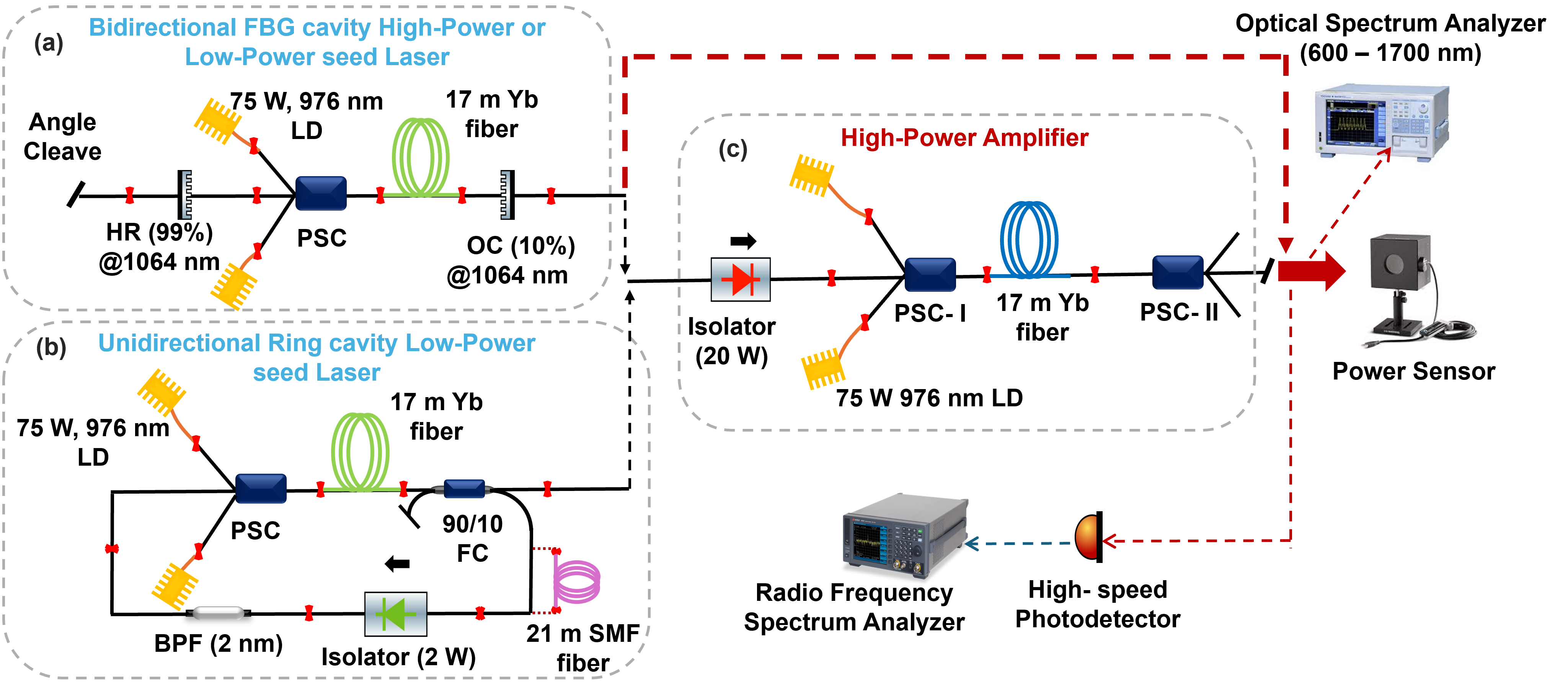}
\vspace{-0.2cm}
\caption{Schematic of the experimental setup. (a)~Bidirectional FBG linear-cavity Yb fiber laser, used as a high-power reference source and, at low power, as an FBG seed. (b)~Unidirectional ring-cavity Yb seed laser. (c)~High-power amplifier seeded by either the FBG or the ring seed laser and operated in forward, backward, or bidirectional pumping. The output is tapped for measurement of the optical spectrum, output power, and RIN. LD: laser diode; PSC: pump-and-signal combiner; FBG: fiber Bragg grating; ISO: isolator; BPF: band-pass filter; OC: optical coupler; Yb DCF: Yb-doped double-clad fiber; PD: photodetector; OSA: optical spectrum analyzer; RSA: RF spectrum analyzer.}
\label{fig:setup}
\end{figure*}

Fig.~\ref{fig:setup} shows the three conventional CW Yb-doped fiber laser sources investigated in this work: a standalone bidirectional FBG oscillator, an FBG-seeded MOPA, and a unidirectional ring-cavity-seeded MOPA. All sources operate near 1064~nm with a $\sim$2~nm 3-dB bandwidth at full power and are cladding-pumped at 976~nm. The oscillator and amplifier stages use the same type of 5/130~$\mu$m Yb-doped double-clad fiber (Coherent), with a gain-fiber length of 17~m, so that differences in RIN arise primarily from the seed-cavity architecture and operating conditions rather than from changes in the gain medium. The first configuration is a bidirectional Yb fiber oscillator formed by a high-reflectivity FBG ($\sim$99\%, 1064~nm, $\sim$2~nm) and a low-reflectivity output-coupler FBG ($\sim$10\%, $\sim$0.5~nm), pumped through a pump-and-signal combiner (PSC). Because the linear cavity supports counter-propagating waves, its longitudinal-mode spacing (FSR) is $\sim$5~MHz. In the ring configuration, the two FBGs are replaced by a closed unidirectional loop of comparable physical length. A 90/10 coupler extracts 90\% of the power and recirculates 10\% through a fiber isolator, which enforces unidirectional operation and suppresses feedback instabilities, and through a 1064~nm band-pass filter, which provides wavelength selection and ASE rejection, before returning to the PSC. Because unidirectional operation gives a mode spacing of twice that of the linear cavity, the ring FSR is $\sim$11.5~MHz. To isolate the role of FSR, the ring cavity was also lengthened by inserting 21~m of 1060XP passive fiber into the loop, reducing the FSR to $\sim$5~MHz, close to that of the FBG cavity.

For the MOPA configurations, the output of either the FBG seed or the ring seed is launched into the same high-power amplifier, which is pumped by two 976~nm laser diodes, each capable of delivering up to 75~W at 12~A. This allows the two seed architectures to be compared using an identical amplifier stage. The amplifier output end is angle-cleaved to suppress parasitic back-reflections. A fiber isolator placed between the seed and the amplifier protects the seed source from backward-propagating reflections and residual pump light. The 20~W power-handling limit of this isolator sets the maximum usable seed power; accordingly, seed powers of 5, 10, and 15~W are investigated. To maintain the same electrical and thermal technical-noise environment across the two seed architectures, the FBG and ring seed lasers were driven using the same pump diode and driver modules, operated at low current when used for seeding, and were amplified using the same amplifier. We found this to be essential: when the seeds were driven by different diodes and drivers, the low-frequency (10--159~kHz) RIN differed between the two architectures, whereas with the same pump sources and drivers, the low-frequency RIN became almost the same for both, indicating that this band is dominated mainly by pump and driver technical noise rather than the cavity architecture. The amplifier pumping geometry is set to forward (both diodes at the input), backward (both at the output), or bidirectional (one diode at each end). These variables--seed-cavity topology, seed power (5--15~W), pumping geometry, and ring FSR ($\sim$11.5 vs.\ $\sim$5~MHz)--define the design space over which the RIN is mapped in this work. For a like-for-like comparison, all architectures are evaluated at a common output power of 72~W, with seed power and pumping geometry matched between the FBG- and ring-seeded systems.

The optical spectrum and output power from each configuration were measured using an optical spectrum analyzer (Yokogawa AQ6370D) and a thermopile power sensor (Gentec), respectively. For the RIN measurements, a small fraction of the output was tapped with a single-mode fiber pickup and coupled to a 30~GHz InGaAs photodetector (Thorlabs DXM30AF). The incident power was kept below 1~mW, giving a DC photocurrent below 700~$\mu$A, well within the detector's linear range of 10~mA maximum. The optical power was adjusted such that the measured noise exceeded the detector/analyzer noise floor by at least $\sim$10~dB wherever permitted by the source noise. The photocurrent noise spectrum was acquired using an RF spectrum analyzer (Keysight N9020B) over three RF bands: 10--159~kHz, 160~kHz--2~MHz, and 2~MHz--1~GHz. The corresponding resolution bandwidths were 1.3~kHz, 18~kHz, and 390~kHz, respectively. For each band, the analyzer noise floor was measured under identical settings with no light incident on the photodetector. Both the measured RF noise and the noise floor were converted to linear power units, and the noise floor was subtracted from the measured spectrum. The resulting noise power was divided by the resolution bandwidth to obtain the noise power spectral density, and then normalized by the DC electrical power $I_{\mathrm{dc}}^{2}\cdot 50~\Omega$, where $I_{\mathrm{dc}}$ is the DC photocurrent and 50~$\Omega$ is the RF input impedance, to obtain the RIN power spectral density. The RIN PSD is reported in dBc/Hz as $10\log_{10}$ of this quantity. The RMS RIN in each RF band was obtained by integrating the linear RIN PSD over the corresponding bandwidth and taking the square root; the resulting values are reported as percentages. The reported integrated RMS RIN values were reproducible across repeated measurements under nominally identical conditions. In the high-frequency window the integrated values were reproducible to within $\sim$2\%, while the low- and mid-frequency values, being dominated by technical noise, showed somewhat larger but consistent run-to-run variation that did not affect the architecture- and band-dependent trends reported here.

\section{Experimental Results}
\subsection{Optical Power and Spectral Characteristics}

\begin{figure}[!t]
\centering
\includegraphics[width=3.5in]{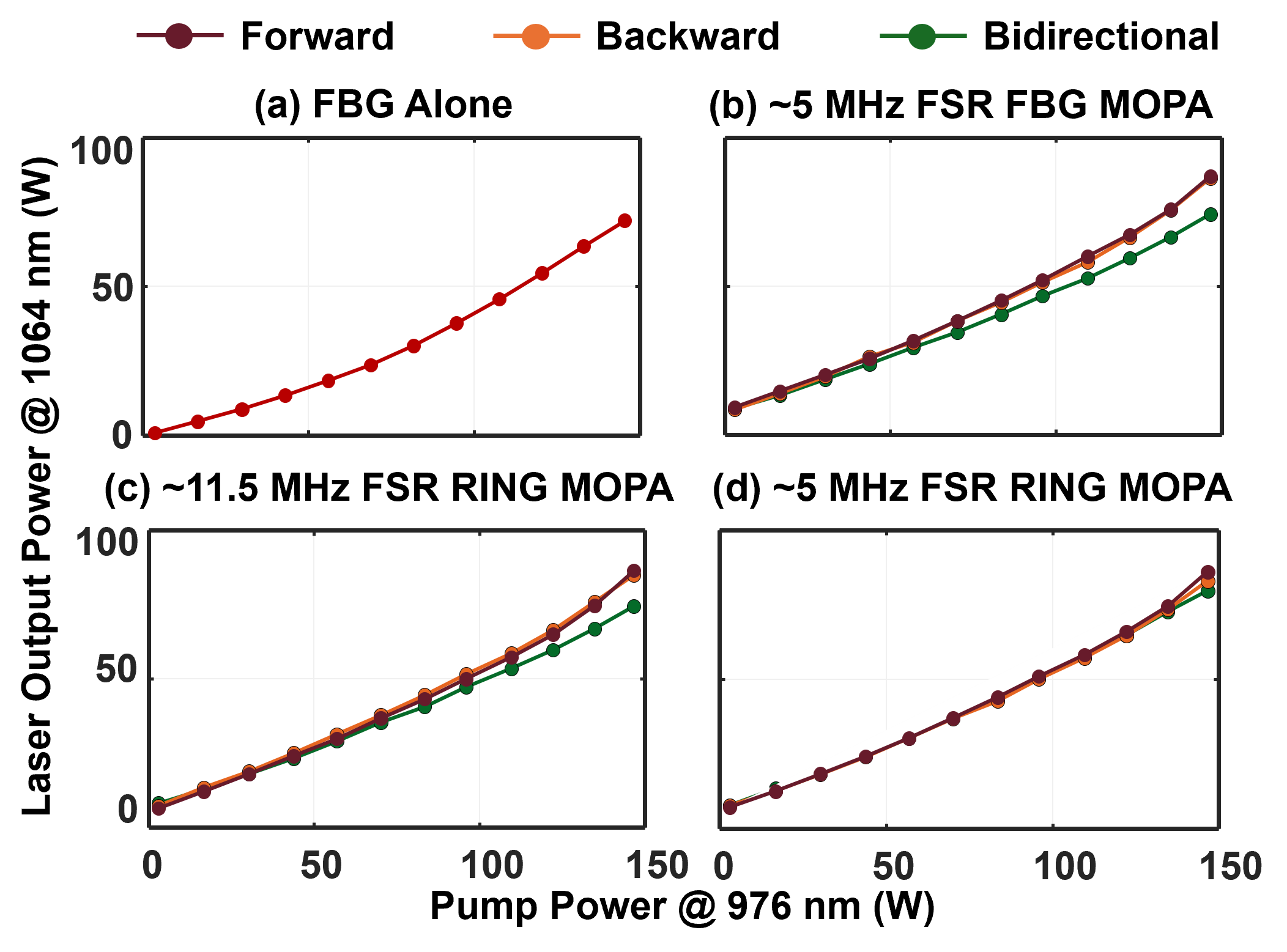}
\vspace{-0.2cm}
\caption{Output power versus pump power for all Yb fiber laser systems; MOPAs under forward, backward, and bidirectional pumping. MOPA: master oscillator power amplifier; FSR: free spectral range.}
\label{fig:outputpower}
\end{figure}

Fig.~\ref{fig:outputpower} shows the output power as a function of pump power for the standalone FBG oscillator and the three MOPA configurations. The MOPA measurements were performed under forward, backward, and bidirectional pumping at a fixed seed power of 15~W. In all cases, the output power increases monotonically with pump power. The standalone FBG oscillator delivers a maximum output power of 72~W at a combined pump power of 145~W. For the MOPA configurations, forward pumping provides the highest output power among the three pumping geometries. At the same maximum pump power, the FBG-seeded MOPA reaches 87~W, while the ring-seeded MOPA with an FSR of $\sim$11.5~MHz reaches 86.5~W. When the ring-cavity FSR is reduced to $\sim$5~MHz by lengthening the loop, the ring-seeded MOPA reaches 86~W under forward pumping at the same seed power. The small differences in output power among the MOPA configurations likely arise from differences in passive loss, seed-cavity spectral characteristics, and pump/gain distribution. Although the standalone FBG oscillator can provide comparable power at the output-coupler FBG, the usable output power at the measurement port is reduced to 72~W because of the additional splice loss associated with the output fiber extension. Therefore, all subsequent spectral and RIN comparisons are performed at a common output power of 72~W.

\begin{figure}[!t]
\centering
\includegraphics[width=3.5in]{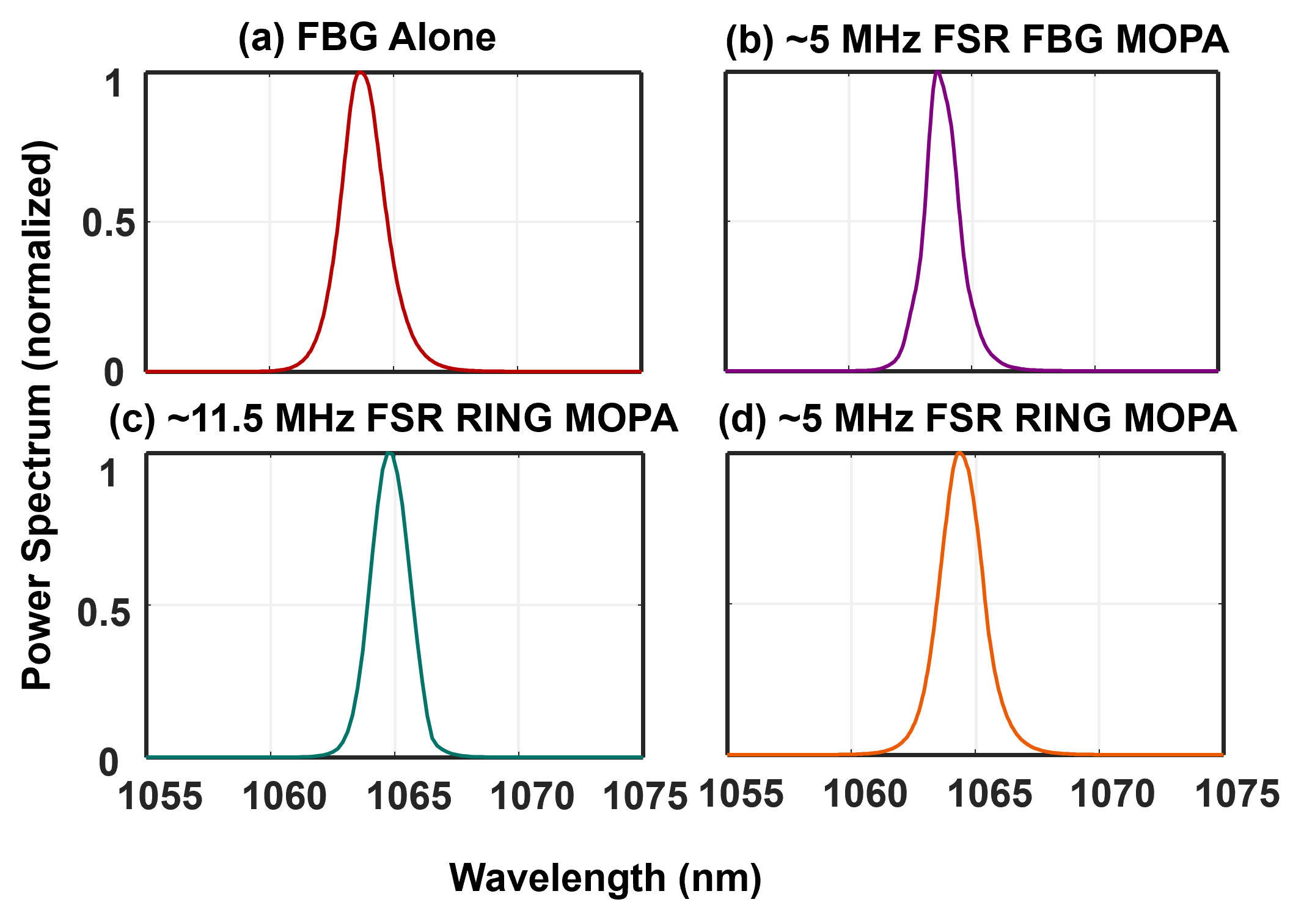}
\vspace{-0.2cm}
\caption{Optical spectra at a common output power of 72~W for all Yb fiber laser systems; MOPAs under forward pumping. MOPA: master oscillator power amplifier; FSR: free spectral range.}
\label{fig:outputspectrum}
\end{figure}

Fig.~\ref{fig:outputspectrum} shows the optical spectra of the standalone FBG oscillator, the FBG-seeded MOPA, and the ring-seeded MOPAs at the common output power of 72~W under forward pumping. All configurations operate near 1064~nm and exhibit comparable spectral bandwidths of approximately 2~nm. No pronounced residual pump component or ASE pedestal is observed within the measured spectral range, indicating efficient pump absorption and comparable spectral quality across the architectures. Therefore, the subsequent RIN differences can be primarily attributed to the seed-cavity topology, seed power, pumping geometry, and ring-cavity FSR rather than to large differences in output power or optical spectrum.
To facilitate a direct comparison between the investigated laser architectures, the RIN spectra and RMS RIN values were analyzed at a common output power of 72~W. This operating point was selected because it was achievable across all configurations while remaining close to their high-power operating regime, thereby allowing the impact of cavity design and pumping configuration on the noise performance to be isolated. 

\begin{figure*}[t]
\centering
\includegraphics[width=\textwidth]{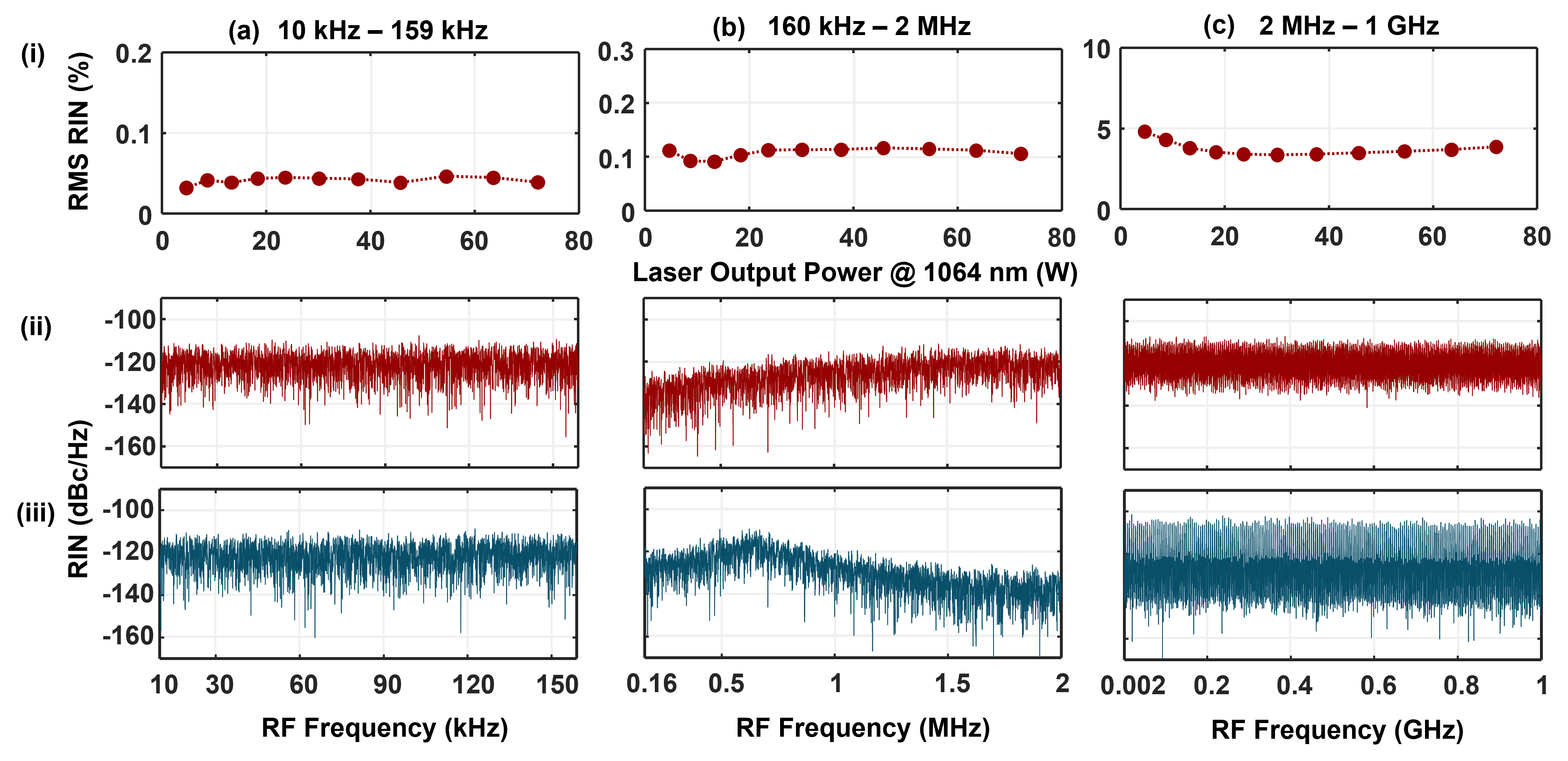}
\vspace{-0.2cm}
\caption{RF-resolved RIN characteristics of the standalone bidirectional FBG oscillator. (i)~Integrated RMS RIN versus output power in the (a)~low-frequency window, 10--159~kHz; (b)~mid-frequency window, 160~kHz--2~MHz; and (c)~high-frequency window, 2~MHz--1~GHz. (ii)~RIN spectra at 72~W output power and (iii)~at 15~W output power, where the FBG oscillator also serves as the seed for the FBG-seeded MOPA.}
\label{fig:fbgalone}
\end{figure*}
\subsection{RIN Characteristics of the Standalone FBG Oscillator}

Fig.~\ref{fig:fbgalone} shows the RF-resolved RIN characteristics of the standalone bidirectional FBG oscillator, which serves as the baseline conventional multi-longitudinal-mode (MLM) source. At 72~W, the integrated RMS RIN values are 0.043\%, 0.106\%, and 3.863\% in the low-, mid-, and high-frequency windows, respectively. The much larger RMS RIN in the high-frequency window does not arise from a proportionally larger RIN spectral density; the band-averaged RIN PSD values remain of the same order across the three windows, around $\sim$--118 to --122~dBc/Hz. Instead, the high-frequency RMS value is dominated by the much wider integration bandwidth of the 2~MHz--1~GHz window. This illustrates why a single integrated RIN number is insufficient: it would be dominated by the broadest RF band and would obscure the distinct noise features relevant to different applications.

The low-frequency RIN remains nearly constant with output power, indicating that this band is mainly limited by technical noise such as pump-driver current fluctuations, thermal drift, and environmental perturbations. This is consistent with the shared pump-source control discussed in Section~\ref{sec:setup}, where using the same pump diode and driver made the low-frequency RIN comparable across seed architectures. 
In contrast, the mid-frequency spectrum contains a relaxation-oscillation feature that shifts with output power. At 15~W, this feature appears near $\sim$0.6~MHz [Fig.~\ref{fig:fbgalone}(iii)(b)], whereas at 72~W it shifts toward the upper edge of the 160~kHz--2~MHz window [Fig.~\ref{fig:fbgalone}(ii)(b)]. This power-dependent shift is consistent with the dependence of relaxation dynamics on gain, intracavity power, and cavity photon lifetime \cite{cheng2020relative}.

The high-frequency spectrum contains pronounced longitudinal-mode beat-note features associated with the $\sim$5~MHz-FSR FBG cavity. At 15~W, these beat notes are well resolved as discrete peaks above a relatively low inter-peak baseline [Fig.~\ref{fig:fbgalone}(iii)(c)]. At 72~W, the beat-note contrast is strongly reduced: the inter-peak pedestal rises and partially fills the gaps between adjacent peaks, producing a denser elevated high-frequency noise floor [Fig.~\ref{fig:fbgalone}(ii)(c)]. This indicates that power scaling in the standalone FBG oscillator redistributes the high-frequency intensity noise from isolated mode-beating features toward a more continuous broadband pedestal. Such redistribution is likely associated with stronger gain saturation, mode competition, nonlinear phase perturbations, and thermally induced cavity fluctuations at higher intracavity power. As a result, the integrated high-frequency RMS RIN can remain of the same order even though the spectral morphology changes substantially.

Since the 15~W FBG oscillator output also serves as the seed for the FBG-seeded MOPA, this measurement defines the seed-noise structure whose transfer through the amplifier--and in particular whether amplification suppresses or enhances this high-frequency pedestal--is examined in Section~\ref{sec:seedpower}.

\subsection{Effect of Seed Power}
\label{sec:seedpower}
\begin{figure*}[t]
\centering
\includegraphics[width=\textwidth]{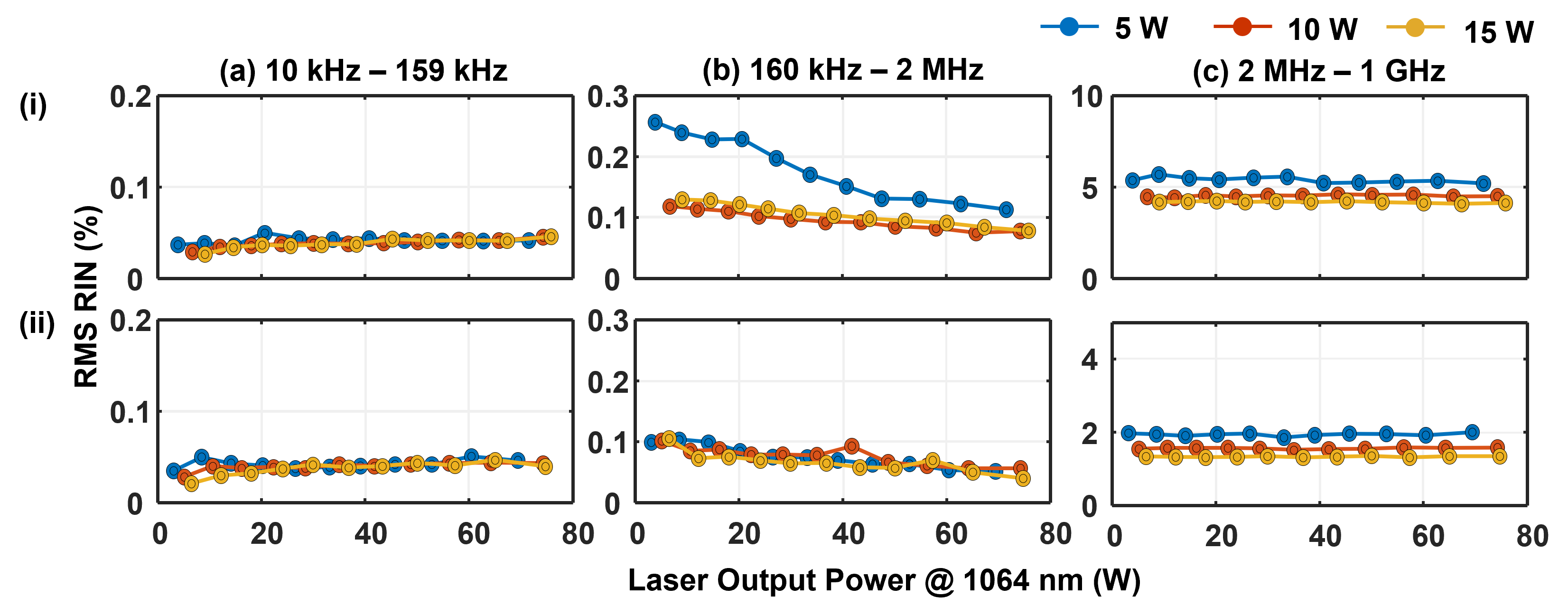}
\vspace{-0.2cm}
\caption{RMS RIN (\%) versus output power for the (i)~FBG-seeded MOPA with $\sim$5~MHz seed-cavity FSR and (ii)~ring-seeded MOPA with $\sim$11.5~MHz seed-cavity FSR at seed powers of 5, 10, and 15~W under forward pumping, in the (a)~low-frequency window, 10--159~kHz; (b)~mid-frequency window, 160~kHz--2~MHz; and (c)~high-frequency window, 2~MHz--1~GHz.}
\label{fig:seedpower}
\end{figure*}

Fig.~\ref{fig:seedpower} shows the effect of seed power on the RF-resolved RMS RIN of the FBG- and ring-seeded MOPAs under forward pumping. The seed power was varied from 5 to 15~W while keeping the pumping geometry fixed, allowing seed power to be isolated as an amplifier operating parameter. The response is strongly RF-band dependent. In the low-frequency window [Fig.~\ref{fig:seedpower}(a)], both architectures show only weak seed-power dependence, with RMS RIN remaining close to $\sim$0.04--0.05\% at the high-power operating point. This confirms that the 10--159~kHz band is dominated mainly by common technical-noise contributions, such as pump-driver fluctuations, thermal drift, and environmental perturbations, rather than by seed-power-dependent amplifier dynamics.

The seed-power dependence becomes more evident above 160~kHz. In the mid-frequency window [Fig.~\ref{fig:seedpower}(b)], increasing the seed power suppresses the RIN relative to the 5~W case, although the trend is not strictly monotonic at the intermediate seed power. For the FBG-seeded MOPA, the RMS RIN decreases from 0.113\% at 5~W seed power to $\sim$0.078\% at 10--15~W. For the ring-seeded MOPA, the lowest value is obtained at 15~W, where the RMS RIN reaches 0.0403\%. This behavior is qualitatively consistent with the frequency-dependent noise-transfer picture established for single-frequency and ultrafast Yb fiber amplifier systems, where gain saturation and inversion dynamics can damp slow seed-power fluctuations, while higher-frequency seed fluctuations are transferred more efficiently through the amplifier \cite{gierschke2019relative,liu2023intensity}. However, the present systems operate in the multi-longitudinal-mode CW regime, where the seed-cavity mode structure introduces additional high-RF noise features. Therefore, the effect of seed power must be interpreted together with the seed-cavity topology. This seed-power suppression is observed for both the seed-alone outputs and the amplified MOPA outputs, as detailed below.

The clearest seed-power effect is observed in the high-frequency window [Fig.~\ref{fig:seedpower}(c)]. For the FBG-seeded MOPA, the RMS RIN decreases from 5.20\% at 5~W seed power to 4.49\% at 10~W and 4.14\% at 15~W. For the ring-seeded MOPA, it decreases from 2.01\% to 1.58\% and 1.35\%, respectively. Thus, higher seed power is most effective in the 2~MHz--1~GHz band. Nevertheless, seed power does not change the ordering of the architectures: at every seed power, the ring-seeded MOPA remains substantially quieter than the FBG-seeded MOPA, giving about 3$\times$ lower high-frequency RMS RIN and about 9--10~dB lower band-averaged RIN PSD.

Comparison with the seed-alone RIN clarifies the origin of this improvement. For the FBG seed, the high-frequency seed RIN varies only modestly from 5 to 15~W, decreasing from about 4.5\% to 3.66\%. Therefore, the reduction in FBG-MOPA output RIN with increasing seed power mainly reflects the amplifier operating deeper in saturation, with only a modest contribution from improved seed RIN. However, the amplified high-frequency RIN remains higher than the corresponding FBG-seed RIN at every seed power; for example, at 15~W seed power, it increases from 3.66\% for the seed to 4.14\% after amplification. This indicates that the FBG-seeded amplifier does not provide net suppression of the high-RF mode-beating-dominated noise and can add or redistribute noise in this band.

The ring-seeded MOPA behaves differently. The high-frequency RIN of the ring seed itself decreases strongly with seed power, from about 6.11\% at 5~W to 1.63\% at 10~W and 1.47\% at 15~W. In addition, the amplified output remains at or below the corresponding ring-seed RIN in all three RF windows; at 15~W, the high-frequency RMS RIN decreases from 1.47\% for the seed to 1.35\% after amplification. Thus, the ring-seeded MOPA benefits from two effects simultaneously: improved seed RIN at higher seed power and saturated-amplifier damping of residual intensity fluctuations. This net broadband RIN reduction is not observed for the FBG-seeded MOPA because its high-frequency mode-beating signature is retained and partly enhanced after amplification.
\begin{figure*}[t]
\centering
\includegraphics[width=\textwidth]{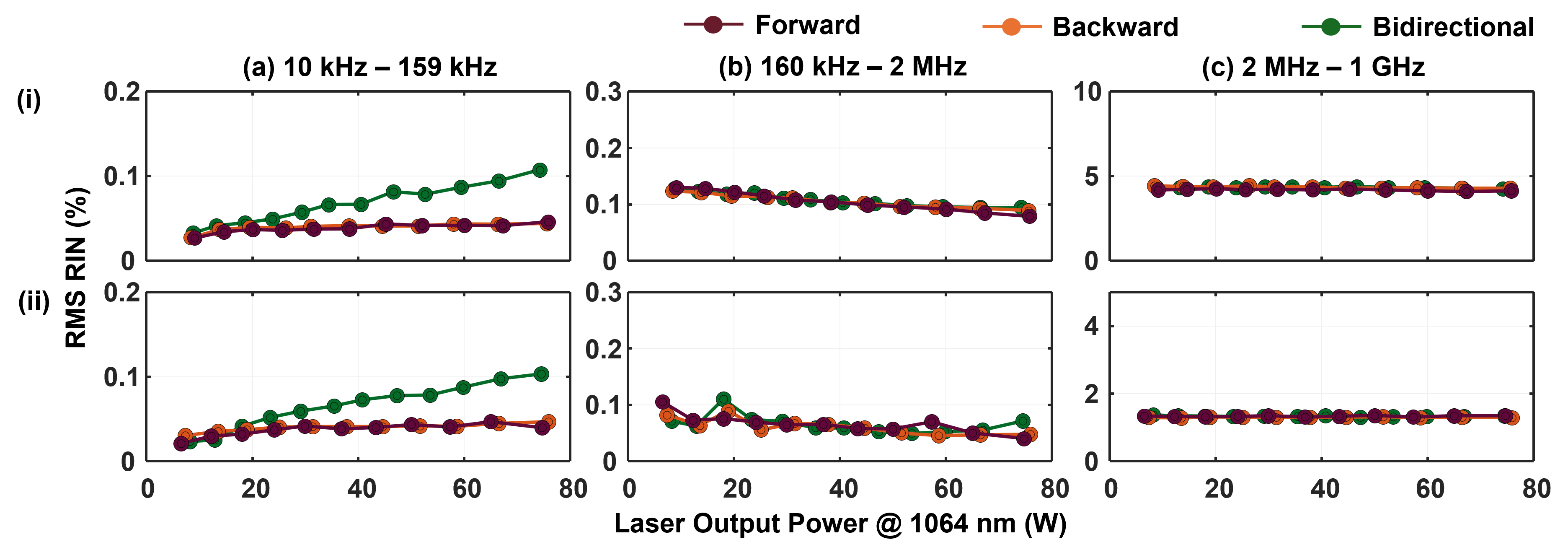}
\vspace{-0.3cm}
\caption{RMS RIN (\%) versus output power under forward, backward, and bidirectional pumping for the (i)~FBG-seeded MOPA with $\sim$5~MHz seed-cavity FSR and (ii)~ring-seeded MOPA with $\sim$11.5~MHz seed-cavity FSR, in the (a)~low-frequency window, 10--159~kHz; (b)~mid-frequency window, 160~kHz--2~MHz; and (c)~high-frequency window, 2~MHz--1~GHz, at a fixed seed power of 15~W.}

\label{fig:pumpcompare}
\end{figure*}
These observations show that seed power is an RF-band-selective RIN-control knob, but its usefulness is architecture-dependent. Increasing seed power has little effect on the technical-noise-limited low-frequency band, provides moderate suppression in the mid-frequency band, and gives the largest improvement in the high-frequency band. More importantly, the amplifier output preserves the RF spectral signature of the seed cavity, so seed power alone cannot compensate for an unfavorable seed-cavity topology. Based on the lowest or near-lowest RIN across the mid- and high-frequency windows, a seed power of 15~W is used for the subsequent pumping-geometry and cavity-FSR comparisons.

\subsection{Effect of Pumping Geometry}
\begin{figure*}[t]
\centering
\includegraphics[width=\textwidth]{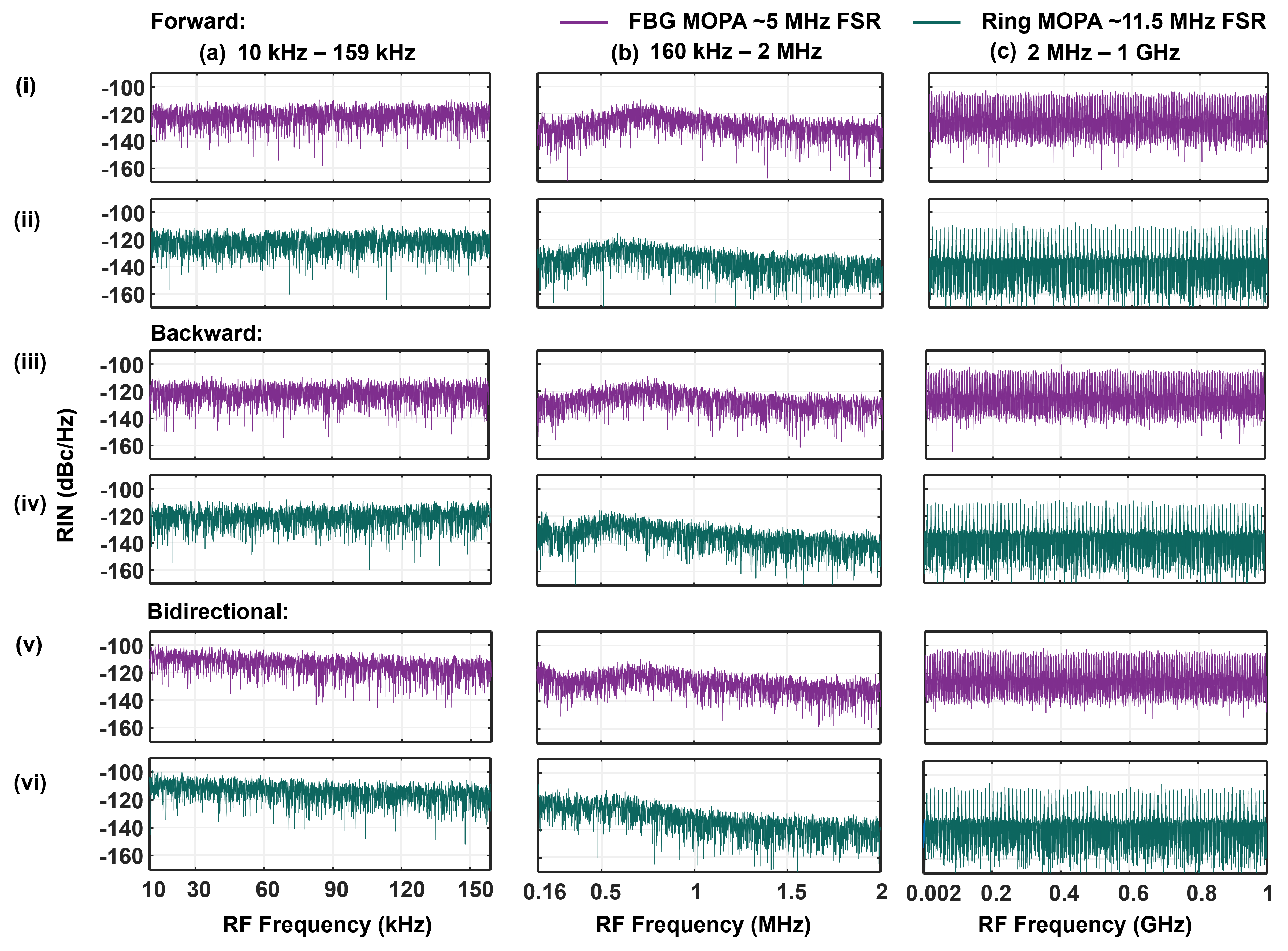}
\vspace{-0.16cm}
\caption{RIN spectra at 72~W output power and 15~W seed power for the FBG-seeded MOPA under (i)~forward, (iii)~backward, and (v)~bidirectional pumping, and for the ring-seeded MOPA under (ii)~forward, (iv)~backward, and (vi)~bidirectional pumping. Columns (a), (b), and (c) correspond to the 10--159~kHz, 160~kHz--2~MHz, and 2~MHz--1~GHz RF windows, respectively.}

\label{fig:pumpspectrumcompare}
\end{figure*}
After fixing the seed power at 15~W, the amplifier pumping geometry was varied to isolate the effect of pump launch direction. In the forward and backward cases, both pump diodes were coupled through the pump-and-signal combiner at one end of the amplifier, whereas in the bidirectional case one pump diode was coupled from each end. The same pump diodes and driver were used in all cases, so the observed differences mainly reflect changes in pump-launch geometry, longitudinal pump/gain distribution, and effective amplifier saturation.

Fig.~\ref{fig:pumpcompare} shows that forward and backward pumping give nearly identical RIN for both MOPA architectures, whereas bidirectional pumping provides no RIN advantage under the present implementation. The most pronounced difference occurs in the low-frequency window [Fig.~\ref{fig:pumpcompare}(a)]. For the FBG-seeded MOPA, the RMS RIN at 72~W is 0.046\% under forward pumping and 0.0437\% under backward pumping, but increases to 0.107\% under bidirectional pumping. A similar trend is observed for the ring-seeded MOPA, where the corresponding values are 0.0399\%, 0.047\%, and 0.103\%, respectively. Since this increase occurs for both seed architectures, it is not caused by the seed-cavity topology. Instead, it points to a pump-configuration-dependent technical-noise contribution. In a saturated fiber amplifier, slow intensity fluctuations are governed by the coupled pump--gain--signal dynamics. Previous studies on Yb fiber amplifiers have shown that pump-noise transfer behaves predominantly as a low-pass process, whereas seed/signal intensity fluctuations are more strongly damped at low Fourier frequencies by gain saturation and inversion dynamics \cite{gierschke2019relative,liu2023intensity}. Therefore, the low-frequency RIN is expected to be especially sensitive to the effective saturation and longitudinal pump/gain distribution of the amplifier. In the present experiment, the bidirectional configuration exhibits lower power-scaling efficiency [Fig.~\ref{fig:outputpower}], indicating a less favorable effective pump/gain distribution and weaker saturation than the forward and backward cases. This can make the output more sensitive to low-frequency pump/gain fluctuations, thereby raising the 10--159~kHz RIN. Additional implementation-specific factors, such as imbalance between the two pump paths, bend-dependent cladding-pump propagation in the coiled gain fiber, and local thermal loading, may also contribute. 

In the mid-frequency window [Fig.~\ref{fig:pumpcompare}(b)], the differences between pumping geometries are smaller but follow the same tendency. For the FBG-seeded MOPA, the RMS RIN changes from 0.078\% under forward pumping to 0.089\% under backward pumping and 0.094\% under bidirectional pumping. For the ring-seeded MOPA, it changes from 0.0403\% to 0.047\% and 0.0556\%, respectively. The spectra in Fig.~\ref{fig:pumpspectrumcompare}(b) show that bidirectional pumping also slightly reshapes the relaxation-oscillation/gain-dynamics feature in the 160~kHz--2~MHz band. This indicates that the longitudinal pump distribution can modify the mid-frequency gain-dynamics response. However, the effect remains much smaller than the architecture-dependent difference between the FBG- and ring-seeded MOPAs. This behavior is qualitatively consistent with saturated-amplifier noise-transfer studies, where gain saturation and inversion dynamics damp slow intensity fluctuations, while higher-frequency seed fluctuations are transferred more efficiently through the amplifier \cite{gierschke2019relative,liu2023intensity}. 

In contrast, the high-frequency window [Fig.~\ref{fig:pumpcompare}(c)] is nearly insensitive to pumping geometry. For the FBG-seeded MOPA, the RMS RIN remains within 4.1--4.3\% across forward, backward, and bidirectional pumping. For the ring-seeded MOPA, it remains within 1.31--1.35\%. The corresponding band-averaged RIN PSD is also nearly unchanged with pump direction, remaining around $-117.4$ to $-117.8$~dBc/Hz for the FBG-seeded MOPA and around $-127.4$ to $-127.6$~dBc/Hz for the ring-seeded MOPA. Therefore, the 2~MHz--1~GHz RIN is governed primarily by the seed-cavity RF signature and its transfer through the amplifier, rather than by the amplifier pumping direction.

The spectra in Fig.~\ref{fig:pumpspectrumcompare} further support this conclusion. Changing the pumping geometry modifies the low-frequency noise floor and slightly alters the mid-frequency relaxation/gain-dynamics feature, especially in the bidirectional case, but it does not remove the characteristic high-frequency RF signature of each seed cavity. The FBG-seeded MOPA retains the denser high-frequency noise structure associated with the $\sim$5~MHz FBG seed cavity, whereas the ring-seeded MOPA retains the lower-noise RF structure associated with the unidirectional ring seed. Thus, pumping geometry is a secondary design knob compared with seed-cavity topology: it can influence technical and gain-dynamics-related noise, but it cannot compensate for an unfavorable high-frequency seed-noise structure. Since forward pumping gives the highest output power while maintaining RIN comparable to or lower than the other pumping geometries, it is used for the subsequent cavity-FSR and architecture comparisons.

\subsection{Effect of Cavity FSR}
\begin{figure*}[t]
\centering
\includegraphics[width=\textwidth]{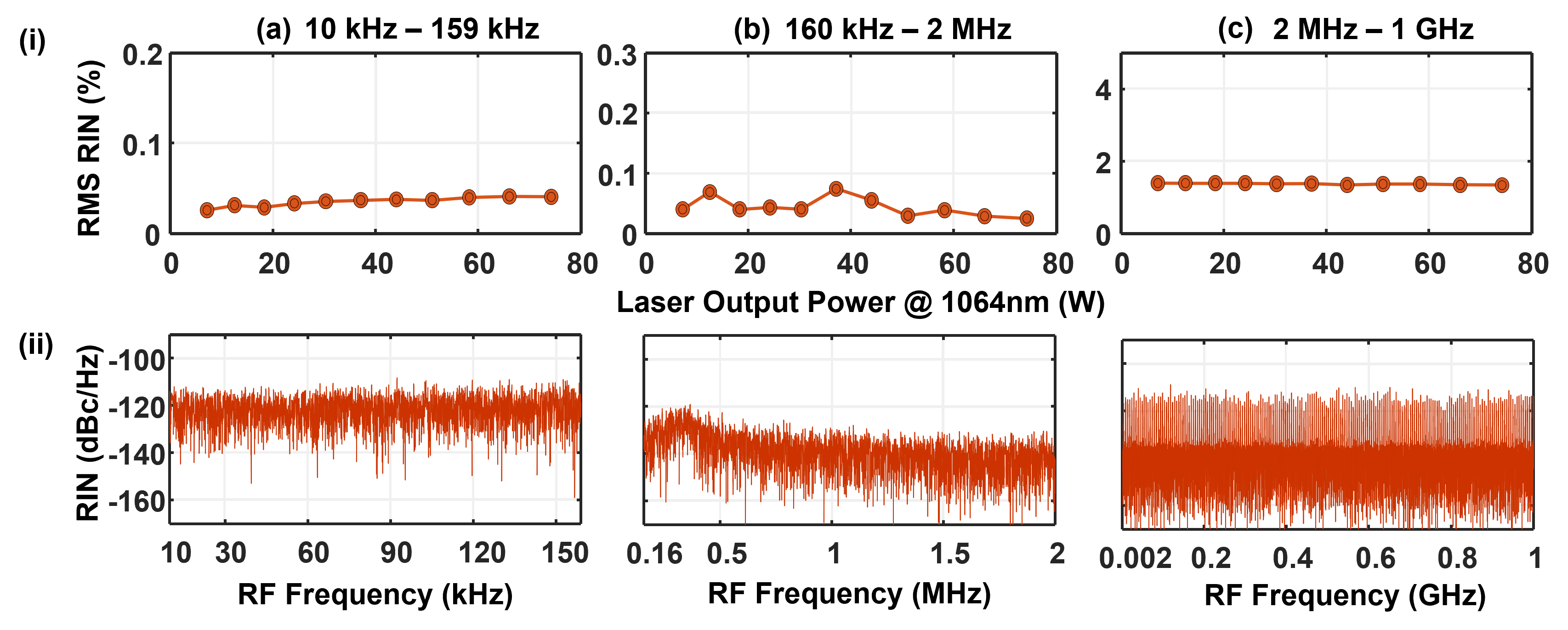}
\vspace{-0.3cm}
\caption{(i)~RMS RIN (\%) versus output power and (ii)~RIN spectra at 72~W output power for the ring-seeded MOPA with $\sim$5~MHz seed-cavity FSR, under forward pumping and 15~W seed power, in the (a)~low-frequency window, 10--159~kHz; (b)~mid-frequency window, 160~kHz--2~MHz; and (c)~high-frequency window, 2~MHz--1~GHz.}
\label{fig:ring5fsr}
\end{figure*}

To examine whether the lower RIN of the ring-seeded MOPA is simply a consequence of its larger longitudinal-mode spacing, the ring seed cavity was lengthened by inserting $\sim$21~m of passive fiber into the loop, reducing its FSR from $\sim$11.5~MHz to $\sim$5~MHz. This reduced-FSR ring seed was then amplified using the same amplifier, 15~W seed power, and forward-pumping configuration used for the original $\sim$11.5~MHz ring-MOPA. Thus, the comparison preserves the unidirectional ring-cavity topology while testing the effect of reducing the cavity FSR to approximately that of the FBG seed cavity.
Fig.~\ref{fig:ring5fsr} shows the RF-resolved RIN of the $\sim$5~MHz-FSR ring-MOPA. At 72~W output power, the RMS RIN values are 0.0408\%, 0.0253\%, and 1.342\% in the low-, mid-, and high-frequency windows, respectively. These values are comparable to those of the original $\sim$11.5~MHz-FSR ring-MOPA, which gives 0.0399\%, 0.0403\%, and 1.347\% under the same seed-power and forward-pumping conditions. Therefore, reducing the ring-cavity FSR from $\sim$11.5~MHz to $\sim$5~MHz does not degrade the high-frequency integrated RIN; instead, the high-frequency RMS RIN remains essentially unchanged, while the mid-frequency RMS RIN is reduced.
The high-frequency window provides the decisive result. The $\sim$5~MHz ring-MOPA gives a high-band RMS RIN of 1.342\%, with a band-averaged RIN PSD of $-127.43$~dBc/Hz, which is nearly identical to the $\sim$11.5~MHz ring-MOPA value of 1.347\% and $-127.40$~dBc/Hz. In contrast, the FBG-seeded MOPA, which also has a seed-cavity FSR of $\sim$5~MHz, gives a much higher high-frequency RMS RIN of 4.14\%, with a band-averaged RIN PSD of $-117.75$~dBc/Hz. The spectra in Fig.~\ref{fig:ring5fsr}(ii)(c) confirm that reducing the ring FSR changes the spacing of the RF beat-note features, as expected, but does not raise the high-frequency noise floor toward the FBG-MOPA level. This shows that the high-frequency RIN advantage of the ring-MOPA is not simply a mode-spacing effect. Rather, it is primarily associated with the unidirectional ring-cavity topology and the corresponding seed RF-noise structure.

In the mid-frequency window, reducing the FSR modifies the relaxation/gain-dynamics response. The RMS RIN decreases from 0.0403\% for the $\sim$11.5~MHz ring-MOPA to 0.0253\% for the $\sim$5~MHz ring-MOPA, with the band-averaged RIN PSD improving from $-130.54$ to $-134.57$~dBc/Hz. Since lengthening the cavity changes not only the mode spacing but also the cavity delay, photon lifetime, and possibly small passive loss, this mid-frequency change should be interpreted as the combined effect of the longer ring cavity rather than of FSR alone. The low-frequency RIN remains essentially unchanged near $\sim$0.04\%, consistent with its technical-noise-dominated origin.

Overall, cavity FSR is a secondary design parameter compared with seed-cavity topology. Reducing the ring FSR changes the RF spacing and redistributes some noise in the mid-frequency gain-dynamics band, but it does not degrade the high-frequency RIN or remove the low-noise advantage of the ring-seeded MOPA. Therefore, the practical design rule is that the seed-cavity topology should first be chosen to suppress high-frequency RIN, while the cavity FSR can then be adjusted to tune the detailed RF spectral structure without substantially increasing the integrated RIN.

\subsection{Comparison of All Architectures}

\begin{figure*}[t]
\centering
\includegraphics[width=\textwidth]{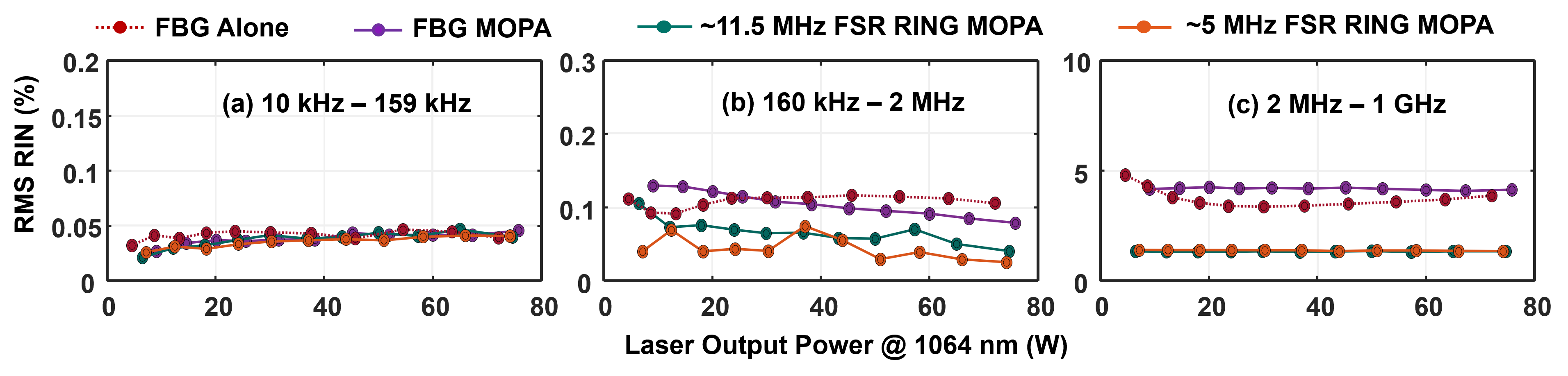}
\vspace{-0.3cm}
\caption{Comparison of RMS RIN (\%) versus output power for the bidirectional FBG oscillator, the FBG-seeded MOPA, and the unidirectional ring-seeded MOPAs with $\sim$11.5~MHz and $\sim$5~MHz FSR, in the (a)~low-frequency window, 10~kHz--159~kHz; (b)~mid-frequency window, 160~kHz--2~MHz; and (c)~high-frequency window, 2~MHz--1~GHz. The MOPA systems are compared at 15~W seed power under forward pumping.}
\label{fig:allconfig}
\end{figure*}

\begin{table*}[t]
\centering
\caption{Integrated RMS RIN (\%) and band-averaged RIN PSD (dBc/Hz) measured at 72~W output power for the investigated CW Yb-doped fiber laser architectures. Values are given as RMS RIN / band-averaged RIN PSD.}
\label{tab:rms_rin}

\renewcommand{\arraystretch}{1.25}
\setlength{\tabcolsep}{4pt}

\begin{tabularx}{\textwidth}{|C|C|C|C|C|}
\hline
\textbf{Frequency Range} &
\textbf{FBG Oscillator} &
\textbf{FBG-seeded MOPA} &
\shortstack{\textbf{Ring-MOPA}\\\textbf{(11.5 MHz FSR)}} &
\shortstack{\textbf{Ring-MOPA}\\\textbf{(5 MHz FSR)}} \\
\hline
10~kHz--159~kHz &
0.043 / $-119.11$ &
0.0457 / $-118.30$ &
0.0399 / $-119.50$ &
0.0408 / $-119.22$ \\
\hline
160~kHz--2~MHz &
0.1058 / $-122.16$ &
0.078 / $-124.70$ &
0.0403 / $-130.54$ &
0.0253 / $-134.57$ \\
\hline
2~MHz--1~GHz &
3.8628 / $-118.25$ &
4.14 / $-117.75$ &
1.3468 / $-127.40$ &
1.3420 / $-127.43$ \\
\hline
\end{tabularx}
\end{table*}

Fig.~\ref{fig:allconfig} and Table~\ref{tab:rms_rin} compare the four investigated architectures: the bidirectional FBG oscillator, the FBG-seeded MOPA, and the unidirectional ring-seeded MOPAs with $\sim$11.5~MHz and $\sim$5~MHz FSR. The MOPA systems are compared at 15~W seed power under forward pumping, which provided the best overall combination of output power and RIN performance. The comparison reveals a clear RF-band-dependent hierarchy.

In the low-frequency window, all four systems exhibit comparable RIN at 72~W output power. The RMS RIN values are 0.0430\%, 0.0457\%, 0.0399\%, and 0.0408\% for the FBG oscillator, FBG-seeded MOPA, $\sim$11.5~MHz-FSR ring-MOPA, and $\sim$5~MHz-FSR ring-MOPA, respectively, with band-averaged RIN PSD levels near $-118$ to $-120$~dBc/Hz. This confirms that the 10~kHz--159~kHz band is primarily limited by common technical-noise sources, such as pump-driver fluctuations, thermal drift, and environmental perturbations. Therefore, this RF band is only weakly affected by seed-cavity topology, amplification, or cavity FSR under the present operating conditions.

The mid-frequency window shows the first clear architecture-dependent separation. The RMS RIN decreases from 0.1058\% for the standalone FBG oscillator to 0.078\% for the FBG-seeded MOPA, indicating that amplification with stronger seed saturation suppresses part of the relaxation/gain-dynamics-related noise. However, replacing the FBG seed with a unidirectional ring seed produces a much larger reduction: the RMS RIN decreases to 0.0403\% for the $\sim$11.5~MHz-FSR ring-MOPA and 0.0253\% for the $\sim$5~MHz-FSR ring-MOPA. In terms of band-averaged RIN PSD, this corresponds to an improvement from $-124.70$~dBc/Hz for the FBG-seeded MOPA to $-130.54$ and $-134.57$~dBc/Hz for the two ring-MOPAs. Thus, the ring-seeded MOPAs are approximately 2--3$\times$ quieter than the FBG-seeded MOPA in RMS RIN, showing that seed-cavity topology plays a larger role than amplification alone in this band. The additional mid-band reduction in the $\sim$5~MHz ring-MOPA is consistent with the longer ring cavity modifying the relaxation/gain-dynamics response.

The high-frequency window gives the most decisive comparison. The two FBG-based systems have comparable high-band RIN: 3.8628\% for the standalone FBG oscillator and 4.14\% for the FBG-seeded MOPA. Thus, amplifying a 15~W FBG seed to 72~W does not suppress the high-RF mode-beating-dominated noise; instead, the FBG-MOPA retains the high-frequency noise signature of the FBG seed cavity. In contrast, both ring-seeded MOPAs exhibit much lower high-frequency RIN, with RMS values of 1.3468\% and 1.3420\% and band-averaged RIN PSD levels near $-127.4$~dBc/Hz. This corresponds to approximately a threefold reduction in RMS RIN and a $\sim$9--10~dB reduction in band-averaged RIN PSD relative to the FBG-seeded MOPA.

Two points from the high-frequency comparison are particularly important. First, the FBG-seeded MOPA remains similar to the standalone FBG oscillator, showing that power amplification does not erase the high-frequency RF signature of a conventional FBG seed. Second, reducing the ring FSR from $\sim$11.5~MHz to $\sim$5~MHz leaves the high-frequency RIN essentially unchanged. Since the $\sim$5~MHz ring-MOPA remains far quieter than the $\sim$5~MHz FBG-seeded MOPA, the ring advantage cannot be explained simply by a larger longitudinal-mode spacing. Instead, the high-frequency RIN is governed primarily by the seed-cavity topology and the resulting RF-noise structure.

This behavior can be understood from the different longitudinal-mode dynamics of the bidirectional FBG and unidirectional ring cavities. In the FBG oscillator, the high-frequency spectrum evolves toward a dense elevated pedestal as the output power increases, indicating that the inter-mode noise floor, not only the discrete beat-note peaks, determines the integrated high-band RIN. The FBG-seeded MOPA inherits this elevated high-frequency noise structure. In contrast, the ring-seeded MOPAs retain discrete RF beat-note features over a lower noise floor, even when the ring FSR is reduced to match the FBG cavity. This suggests that the unidirectional ring cavity suppresses the mechanisms that raise the inter-mode noise floor in the bidirectional FBG cavity, likely by avoiding standing-wave spatial hole burning and reducing feedback-induced mode coupling. A more direct test would compare the same ring cavity with and without the intracavity isolator, isolating the effect of unidirectional operation alone. This was not possible in our system because removing the isolator doubles the intracavity power and exceeds the $\sim$2~W limit of the band-pass filter, restricting bidirectional operation to $\sim$1~W; such a low-power comparison would not be like-for-like, and a full-power test using a higher-power filter is left to future work.

Overall, these results establish a practical RF-resolved design rule for high-power multi-longitudinal-mode CW Yb fiber laser pump sources. Low-frequency RIN is mainly technical-noise limited and should be addressed through pump-driver, thermal, and mechanical stabilization. Mid-frequency RIN can be shaped by amplifier saturation, pump/gain distribution, and cavity length. High-frequency RIN, however, is primarily set by the seed-cavity RF signature and its transfer through the amplifier. Therefore, low MHz--GHz RIN cannot be obtained simply by amplifying a conventional FBG seed; it requires choosing a seed topology that intrinsically provides a lower high-frequency noise floor. In the present comparison, the unidirectional ring-cavity seed provides this advantage while retaining the simplicity, scalability, and practicality of conventional multi-longitudinal-mode CW Yb fiber laser architectures.
\section{Summary and Conclusions}
We have presented a controlled, like-for-like experimental mapping of how four practical design parameters--seed-cavity topology, seed power, amplifier pumping geometry, and seed-cavity free spectral range (FSR)--jointly shape the RF-resolved relative intensity noise (RIN) of high-power MLM CW Yb-doped fiber laser sources from 10~kHz to 1~GHz, using a standalone bidirectional FBG oscillator, an FBG-seeded MOPA, and unidirectional ring-seeded MOPAs with FSRs of $\sim$11.5 and $\sim$5~MHz compared at a common 72~W output.

The RIN separates into three regimes with distinct origins and design levers. The low-frequency band (10--159~kHz) is technical-noise limited and nearly architecture-independent, and is best addressed through pump-driver, thermal, and mechanical stabilization. The mid-frequency band (160~kHz--2~MHz) is shaped by relaxation and gain dynamics, and responds to amplifier saturation, seed power, and cavity length. The high-frequency band (2~MHz--1~GHz) arises from the seed longitudinal-mode structure--discrete mode-beat notes together with the inter-mode noise floor between them--and is governed primarily by the seed-cavity topology and its transfer through the amplifier. A conventional bidirectional FBG seed develops an elevated inter-mode pedestal that is retained, and can even be enhanced, through amplification (high-band RMS RIN rising from 3.66\% for the 15~W seed to 4.14\% at the amplified output), whereas a unidirectional ring seed retains discrete beat notes over a low noise floor and is further damped by the saturated amplifier (1.47\% to 1.35\%). Because the ring advantage persists when the ring FSR is reduced to match that of the FBG cavity, it reflects the unidirectional cavity topology rather than longitudinal-mode spacing.

The practical consequence is that low MHz--GHz pump RIN cannot be obtained simply by amplifying a conventional FBG seed; it requires a seed topology that intrinsically provides a low high-frequency noise floor. Among the systems studied, the $\sim$5~MHz-FSR ring-seeded MOPA under forward pumping with a 15~W seed gave the best overall performance, reducing the high-frequency integrated RMS RIN by $\sim$3$\times$ ($\sim$9--10~dB in band-averaged PSD, to near $-127.4$~dBc/Hz) relative to FBG seeding. These rules are established within the multi-longitudinal-mode, $\sim$2~nm-bandwidth CW regime and the power range studied here, and their extension to substantially different powers, gain fibers, or seed linewidths remains to be verified. Because this route retains the simplicity, low cost, and scalability of conventional MLM CW Yb fiber architectures, it offers a directly deployable path to RF-band-engineered pump sources for supercontinuum generation, cascaded random-Raman lasers, and frequency conversion, where a noise spectrum matched to the target process--rather than a single low RIN figure--determines system performance.


%


\section*{Acknowledgment}

The authors thank Srikar Anumanchipally and Mrittik Chakraborty for their assistance with the experiments. The authors also acknowledge the 5G Laboratory, Department of Electrical Communication Engineering, Indian Institute of Science, Bengaluru, for providing access to the RF spectrum analyzer used in this work.

\section*{Conflict of Interest}
The authors declare no conflicts of interest.

\ifCLASSOPTIONcaptionsoff
  \newpage
\fi

\end{document}